\documentclass[12pt,letterpaper]{article}
\usepackage{t1enc}
\usepackage[latin1]{inputenc}
\usepackage[english]{babel}
\usepackage{graphicx}
\usepackage{amsmath}
\renewcommand{\baselinestretch}{1.75}

\DeclareMathOperator{\atantwo}{atan2}

\begin{document}

\title{Multimodal Symmetric Circular Distributions Based on Nonnegative Trigonometric Sums and a Likelihood Ratio Test for Reflective Symmetry}
\renewcommand{\baselinestretch}{1.00}
\author{Fernández-Durán, J.J. and Gregorio-Domínguez, M.M. \\
ITAM \\
e-mail: jfdez@itam.mx}
\date{}
\maketitle

\begin{abstract}
Fernández-Durán (2004) developed a family of circular distributions based
on nonnegative trigonometric sums (NNTS) which is flexible for modeling datasets exhibiting multimodality and asymmetry. 
Many datasets involving angles in the natural sciences, such as animal movement in biology, are expected to exhibit
reflective symmetry with respect to a central angle (axis) of symmetry. Testing for symmetry in 
the underlying circular density from which these angles are generated is crucial. Additionally, such densities often display multimodality.
This paper identifies the conditions under which NNTS distributions are reflective symmetric and develops a likelihood ratio test for reflective symmetry.
The proposed methodology is demonstrated through applications to simulated and real datasets.
\end{abstract}

\textbf{Keywords}: Symmetry, Nonnegative trigonometric sums, Likelihood Ratio Tests, Maximum Likelihood Estimation, Bootstrap, Asymptotics

\renewcommand{\baselinestretch}{1.75}

\section{Introduction}

Circular data arise in many areas of science, particularly when the measurements are angles or directions. Often, such data exhibit symmetry with respect to a central angle, a feature termed
reflective symmetry. Testing for reflective symmetry in circular distributions is of significant interest, particularly when the data display multimodality.
Not rejecting the null hypothesis of circular reflective symmetry allows the use of more parsimonious circular density functions with important implications for the practical situation and data being modelled. 

In practical situations, the choice of a test for reflective symmetry in circular data depends on the characteristics of the dataset being modeled. Early tests for reflective symmetry
in circular data were adapted from symmetry tests for linear data, such as runs tests and the Wilcoxon signed-rank tests, as reviewed by Pewsey (2002 and 2004). 
One of the first tests specifically developed for circular reflective symmetry was introduced by Schach (1969). This test is a nonparametric rank-based approach to assess symmetry with respect to a given axis, against the alternative such as a displaced or rotated axis of symmetry. Pewsey (2002) proposed an omnibus test for reflective symmetry about an unknown mean direction based on the second central trigonometric sample moment. This test, known as the $\bar{b}_2$ test, uses the test statistic: 
\begin{equation}
\bar{b}_2=\frac{b_2}{\hat{Var}(b_2)}
\end{equation}
with $b_2=\frac{1}{n}\sum_{k=1}^{n}\sin(2(\theta_k-\bar{\theta}))$ and $\bar{\theta}$ being the argument of the first trigonometric moment in its complex numbers form $\bar{R}_1e^{i\bar{\theta}}=\frac{1}{n}\sum_{k=1}^n e^{i\theta_k}$ for a sample of angles $\theta_1, \theta_2, \ldots, \theta_n$. This test statistic follows an asymptotic standard normal distribution and can also be applied using bootstrapping. However, its power has been primarily studied for unimodal circular distributions. As pointed by Pewsey (2002),
the $b_2$ statistic was used by Batschelet (1965) as a coefficient of skewness for circular datasets. 

Pewsey (2004) later proposed the $b_2^*$ test for reflective symmetry about a known median axis. Ley and Verdebout (2014) developed a reflective symmetry test for a fixed center, with alternative hypothesis the k-sine distribution (Abe and Pewsey, 2011). The k-sine distribution is derived by perturbing a base symmetric circular distribution (Umbach and Jammalamadaka, 2009) and is defined as: 
\begin{equation}
f_{\mu,\lambda}^{k^*}(\theta)=f_0(\theta - \mu)[1+\lambda \sin(k^*(\theta-\mu))]
\label{ksine}
\end{equation}
with parameters $\mu$, $\lambda$ and $k^*$ where $f_0(\theta)$ is the base symmetric circular density, and the term $[1+\lambda \sin(k^*(\theta-\mu))]$ introduces perturbation. If $\lambda = 0$, the base symmetric density is recovered. Common models for the base symmetric model include von Mises, cardioid and wrapped Cauchy distributions. When $\lambda \ne 0$, various parameter values for $k^*$ and $\lambda$ allow for densities exhibiting multimodality and skewness.

Ley and Verdebout (2014) introduced a reflective symmetry test with alternative the k-sine model. This test is derived via the Le Cam uniform local asymptotic normality (ULAN) approach (Le Cam, 1986). However, the choice of the $k^*$ value in the $k$-sine alternative model is unclear. Different values of $k^*$ may lead to varying conclusions, as demonstrated in the examples provided
in their paper. A significant limitation of the tests by Ley and Verdebout (2014) and Pewsey (2004) is the requirement for a known axis of symmetry. Ameijeiras-Alonso et al. (2021) extended the test by Ley and Verdebout (2014) accommodating cases where the symmetry axis (central direction) is unknown. The alternative model in this test is the k-sine distribution. They first considered the base symmetric model of the k-sine distribution as known and later addressed scenarios where the base model is unknown, though the parameter $k^*$ is assummed to be known. The application of the Ameijeiras-Alonso et al. (2021) tests for real examples showed, as in the Ley and Verdebout (2014) case, different results depending on the value of the parameter $k^*$.
Note that the $k^*$ parameter determines the maximum number of modes in the k-sine density. Ameijeiras-Alonso et al. (2021) recommended using the Pewsey's (2002) $\bar{b}_2$ test or its bootstraped version when the parameters $\mu$, $k^*$, and the base symmetric model of the $k$-sine distribution are unknown. Note that this is generally the common case in practice.
From a Bayesian perspective, Salvador and Gatto (2022) developed Bayesian tests for circular reflective symmetry of the generalized von Mises model, which can be symmetric or asymmetric but is limited to being unimodal or bimodal. 

Fernández-Durán (2004) derived a family of circular distributions based on nonnegative trigonometric sums. In brief,
the nonnegative trigonometric sum is expressed as the squared norm of a complex number. If $\theta$ is a
circular random variable, $\theta \in (0,2\pi]$, then the circular density function based on
nonnegative trigonometric sums (henceforth referred to as the NNTS density) is expressed as:

\begin{equation}
f(\theta ; M, \underline{c}) = \frac{1}{2\pi}\left|\left|\sum_{k=0}^{M}c_k e^{ik\theta}\right|\right|^2 = \frac{1}{2\pi}\sum_{k=0}^M\sum_{l=0}^M c_k\bar{c}_l e^{i(k-l)\theta}
\label{nntsdensity}
\end{equation}

where $i=\sqrt{-1}$, $c_k$ are complex numbers ($c_k=c_{rk} + ic_{ck}$ for $k=0, \ldots, M$), and $\bar{c}_k=c_{rk} - ic_{ck}$ is the conjugate of $c_k$. To ensure the density function integrates to one, the following constraint on the $c$ parameters must be imposed:
\begin{equation}
\sum_{k=0}^M ||c_k||^2 = 1
\end{equation}
with $c_{c0}=0$ and $c_{r0} \ge 0$, meaning $c_0$ is a non-negative real number. The $\underline{c}$-parameter space corresponds to the surface of a unit hypersphere.
The total number of $c$ free parameters is equal to $2M$, with $M$ being an additional parameter that defines the total number of terms in the sum forming the NNTS density. 
By expressing $c_k=||c_k||e^{i\phi_k}$, where $||c_k||$ is the modulus
of $c_k$ defined as $||c_k||=\sqrt{c_{rk}^2+c_{ck}^2}$, and $\phi_k$ is the argument of $c_k$, $Arg(c_k)$, defined as $\phi_k=Arg(c_k)=\atantwo(c_{ck},c_{rk})$ (using the generalized arctangent function), the NNTS density function can be written as:
\begin{equation}
f(\theta ; M, \underline{c}) = \frac{1}{2\pi}\sum_{k=0}^M\sum_{l=0}^M ||c_k|| ||c_l|| e^{i((k-l)\theta + \phi_k - \phi_l)}.
\label{nntsdensityphiangles}
\end{equation}

The NNTS family of circular distributions is capable of modeling datasets that exhibit skewness and multimodality. Additional properties of NNTS circular models are detailed in Fernández-Durán (2004 and 2007). The main objective of this paper is to identify the conditions for the parameters $c$ under which the NNTS density is reflective
symmetric, i.e., $f(\theta; M, \underline{c}) = f(2\mu - \theta; M, \underline{c})$ for a fixed value $\mu$, referred to as the angle (or axis) of symmetry. Additionally, this work aims to 
develop a parametric likelihood ratio test for reflective symmetry in NNTS distributions. As noted by Pewsey (2002), Kappenman (1988) suggested enhancing the power of nonparametric tests for symmetry in circular data by employing a flexible family of parametric models that include both symmetric and asymmetric distributions. This approach allows the use of parametric likelihood ratio tests to evaluate reflective symmetry. The methodology outlined in this paper aligns with this recommendation by leveraging the NNTS family of models.  
As noted by Mardia (1972, p. 10) \emph{``symmetrical distributions on the circle are comparatively rare''}. However, the NNTS family includes highly flexible symmetrical cases, which can also be multimodal. Additionally, the NNTS family of (symmetric) distributions is nested: NNTS densities with $M=M^*$ are special cases of NNTS models with $M=M^{**}$, where $M^*<M^{**}$. This nesting property is further explored in subsequent sections, where symmetric NNTS models are shown to be particular cases of asymmetric NNTS models.

The symmetric NNTS models are nested provided that models with different values of $M$ share the same angle of symmetry. Because symmetric NNTS models are nested in asymmetric NNTS models and the parameters of the NNTS model are estimated using the method of maximum likelihood (Fernández-Durán and Gregorio-Domínguez, 2010), the application of likelihood ratio tests is straightforward. This allows for the use of all the optimal properties of likelihood ratio tests when testing for circular reflective symmetry in practical applications. 
 
The paper is divided into five sections, including this introduction. The second section outlines the conditions that the NNTS $c$ parameters
must satisfy for the density function to be reflective symmetric and describes the development of the likelihood ratio test for reflective symmetry. 
In the third section, a simulation study is conducted to confirm the size and power of the likelihood ratio test of circular reflective symmetry. The fourth section applies the proposed test to
real datasets. Finally, the conclusions of this study are presented in the fifth section.

\section{Conditions for NNTS Circular Reflective Symmetry and a Likelihood Ratio Test for Circular Reflective Symmetry}

The condition for (reflective) symmetry can be expressed as:
\[
f(\theta; M, \underline{c}) = f(2\mu -\theta; M, \underline{c})
\]
where the parameter $\mu \in (0,2\pi]$ is the angle (axis) of symmetry. It is noteworthy that the general NNTS distribution with $M=1$ is always symmetric.

Without loss of generality, for $M \ge 2$, consider that given $\mu=0$, then the condition for symmetry is:
\[
f(\theta; M, \underline{c}) = f(-\theta; M, \underline{c}).
\]
If $f$ is a member of the NNTS family, then:
\begin{equation}
\frac{1}{2\pi}\sum_{k=0}^M\sum_{l=0}^M c_k\bar{c}_l e^{i(k-l)\theta}=\frac{1}{2\pi}\sum_{k=0}^M\sum_{l=0}^M c_k\bar{c}_l e^{i(l-k)\theta}
\end{equation}
which will be satisfied for all values of $\theta$ if and only if:
\begin{equation}
c_k\bar{c}_l = c_l\bar{c}_k,
\end{equation}
for $k,l=0, \ldots, M$. Given that $c_0$ is a real positive number, this condition is satisfied if $\bar{c}_k=c_k$, 
implying that $c_k$ should be a real number for $k=0, \ldots, M$, or equivalently, the argument of $c_k$, $\phi_k$, should be equal to zero. An NNTS density function where all of the $c$ parameters are real numbers will be reflectively symmetric with angle of symmetry equal to 0. To include an angle of symmetry different from 0, one can work with the variable $\theta - \mu$. The NNTS symmetric density around $\mu$ is expressed as:
\begin{equation}
f(\theta; M, \underline{c}, \mu)= \frac{1}{2\pi}\sum_{k=0}^M\sum_{l=0}^M \rho_k\rho_l e^{i(k-l)(\theta - \mu)}
\end{equation}
with $\rho_k$ real numbers for $k=0, \ldots, M$. Given that $\rho_k$ are real numbers for $k=0, 1, \ldots, M$ and under the restriction that $\sum_{k=0}^M \rho_k^2 = 1$, the total number of free parameters for a symmetric NNTS distribution is $M + 1$. This is because $\rho_0$ can be expressed as $\rho_0=\sqrt{1-\sum_{k=1}^M \rho_k^2}$. The parameters $c_0, c_1, \ldots, c_M$ of the general NNTS density in Equation \ref{nntsdensity} are invariant under translation of the random angle $\theta$ from $\theta$ to $\theta - \mu$ for any angle $\mu$. As a result, the symmetric NNTS density is defined as:
\begin{equation}
f(\theta; M, \underline{c}, \mu)= \frac{1}{2\pi}\sum_{k=0}^M\sum_{l=0}^M ||c_k|| ||\bar{c}_l|| e^{i(k-l)(\theta - \mu)} = 
\frac{1}{2\pi}\sum_{k=0}^M\sum_{l=0}^M c_{Sk} \bar{c}_{Sl} e^{i(k-l)\theta}.
\label{nntsdensitysymmetric}
\end{equation}
Here, $c_{Sk}$ are the $c$ parameters for the symmetric case, satisfying $c_{Sk}=||c_k||e^{-ik\mu}$, with $c_k$ being the parameters of the general NNTS model in Equation \ref{nntsdensity}.
Therefore, the symmetric NNTS model in Equation \ref{nntsdensitysymmetric} is a particular case of the general NNTS model in Equation \ref{nntsdensity} in which the norms (moduli) of the $c$ parameters are
equal in both models, and for the general NNTS model to become a symmetric NNTS model, it is necessary that $\phi_k=-k\mu$ for $k=1,2, \ldots, M$. Given this property,
the maximum likelihood estimation of the parameters of the symmetric NNTS model can be performed by following these steps:

\begin{enumerate}
\item Fit the general NNTS model and obtain the norms of the $c_G$ parameters, $||c_{Gk}||$, for $k=0, 1, \ldots, M$. Use these as initial values for subsequent iterations. 
\item Estimate the angle $\mu$ by maximizing the log-likelihood of a symmetric NNTS model for the translated data, $\theta - \mu$.
\item Re-estimate the parameters $\rho_k$ for $k=0, 1, \ldots, M$ based on the estimate of $\mu$.
\end{enumerate}
These steps are iterated until convergence is achieved in terms of the gradient of the log-likelihood function for the symmetric NNTS model. Computational routines for estimating symmetric NNTS models are available in the $R$ package $CircNNTSR$ (see Fernández-Durán and Gregorio-Domínguez, 2016).

Since the NNTS reflectively symmetric distribution is a special case of the general NNTS distribution (with restrictions on the $c$ parameters), a likelihood ratio
test can be implemented. For a fixed $M$, the likelihood ratio compares the best general NNTS model to the best symmetric NNTS model. Let $l_G(\underline{c},M \mid \underline{\theta})$ represent the log-likelihood for the general NNTS model and $l_S(\underline{c},M \mid \underline{\theta})$ represent the log-likelihood for the symmetric NNTS model. The likelihood ratio statistic is given by:
\begin{equation}
LR_{GS}=-2(l_S(\underline{c},M \mid \underline{\theta}) - l_G(\underline{c},M \mid \underline{\theta})).
\end{equation}
Asymptotically, although the information matrix is singular (see El-Helbawy and Hassan, 1994), $LR_{GS}$ follows a chi-squared distribution with $M-1$ degrees of freedom for $M \ge 2$. This results from imposing $M-1$ constraints $Arg(c_k)=kArg(c_1)$ for $k=2, \ldots, M$ to derive the symmetric NNTS model from the general NNTS model. Note that the general NNTS distribution with $M=1$ is always symmetric and corresponds to the cardioid distribution. The NNTS density with $M=0$ corresponds to the circular uniform density.

The parametric bootstrap implementation of the likelihood ratio test can be performed as suggested by McLachlan (1987). The steps of the bootstrapped likelihood ratio ($BLR$) test are as follows:

\begin{enumerate}
\item Fit the null and alternative models to the original data.
\item Generate $K$ parametric bootstrap replicates, of the same size as the original data, from the null model using parameter estimates from the original data. 
\item For each bootstrap replicate, fit the null and alternative models and compute the test statistic $LR_{GS}$ to obtain $K$ values of the test statistic. 
\item Calculate the p-value of the test as the percentage of bootstrap test statistic values that are greater than or equal to the test statistic obtained from the original dataset.
\end{enumerate}

Alternatively, a Wald type test statistic can be considered. This statistic, $W_{GS}$, is based on the standardized difference between the (general) maximum likelihood estimator, $\hat{\underline{c}}_G$, and the maximum likelihood estimator under the null hypothesis of symmetry, $\hat{\underline{c}}_S$. It is defined as:
\begin{equation}
W_{GS} = (\hat{\underline{c}}_G-\hat{\underline{c}}_S)^H(n(I-\hat{\underline{c}}_G\hat{\underline{c}}_{G}^{H}))(\hat{\underline{c}}_G-\hat{\underline{c}}_S)
\end{equation}
where $n(I-\hat{\underline{c}}_G\hat{\underline{c}}_{G}^{H})$ is the Hessian matrix from the maximum likelihood method (see Fernández-Durán and Gregorio-Domínguez, 2010) and $\hat{\underline{c}}_{G}^{H}$ represents the Hermitian (transpose and conjugate) of $\hat{\underline{c}}_{G}$. The Wald type test statistic, $W_{GS}$, can be simplified to:
\begin{equation}
W_{GS} = n\left(1 - \left(\sum_{k=0}^M ||\hat{c}_{Gk}||^2 \cos(\hat{\phi}_k + k\hat{\mu})\right)^2 - \left(\sum_{k=0}^M ||\hat{c}_{Gk}||^2 \sin(\hat{\phi}_k + k\hat{\mu})\right)^2\right)
\end{equation}
and can finally be expressed as:
\begin{equation}
W_{GS} = n\left(1 - \left|\left|\sum_{k=0}^M \hat{c}_{Gk} ||\hat{c}_{Gk}|| e^{-ik\hat{\mu}}\right|\right|^2\right) = n(1 - ||\hat{\underline{c}}_G^H\hat{\underline{c}}_S||^2 ).
\end{equation}
Here $\hat{c}_{Gk}=||\hat{c}_{Gk}||e^{i\phi_k}$, where $||\hat{c}_{Gk}||$ is the norm (modulus) of $\hat{c}_{Gk}$, and $\hat{\phi}_k=Arg(\hat{c}_{Gk})$ is its argument (angle).
A measure of skewness for NNTS models, $S_{NNTS}$, can also be defined as follows:
\begin{equation}
SK_{NNTS}=\frac{W_{GS}}{n}=1 - ||\hat{\underline{c}}_G^H\hat{\underline{c}}_S||^2.
\end{equation}
The skewness measure $SK_{NNTS}$ takes values in the interval $[0,1]$, with $SK_{NNTS}=0$ for an NNTS reflective symmetric density, and $SK_{NNTS} \ne 0$ indicating asymmetry.

To compare the values of the proposed skewness measure $SK_{NNTS}$, which is nearly zero for symmetric unimodal or multimodal datasets, one can calculate the commonly used sample circular skewness coefficient, $\hat{s}$, defined as (Mardia and Jupp, 2000):
\begin{equation}
\hat{s} = \frac{\bar{R}_2\sin(\bar{\theta}_2-2\bar{\theta}_1)}{(1-\bar{R}_1)^{\frac{3}{2}}}
\end{equation}
where $\bar{R}_1e^{i\bar{\theta}_1}=\frac{1}{n}\sum_{k=1}^n e^{i\theta_k}$ and $\bar{R}_2e^{i\bar{\theta}_2}=\frac{1}{n}\sum_{k=1}^n e^{2i\theta_k}$ are the first and second sample
trigonometric moments in their complex forms for a sample of angles $\theta_1, \theta_2,
\ldots, \theta_n$. As noted by Mardia and Jupp (2000), the sample circular skewness coefficient is nearly zero for symmetric unimodal datasets.

\section{Simulation Study}

\begin{figure}[h]
\center{
\includegraphics[scale=.9, bb= 0 0 504 504]{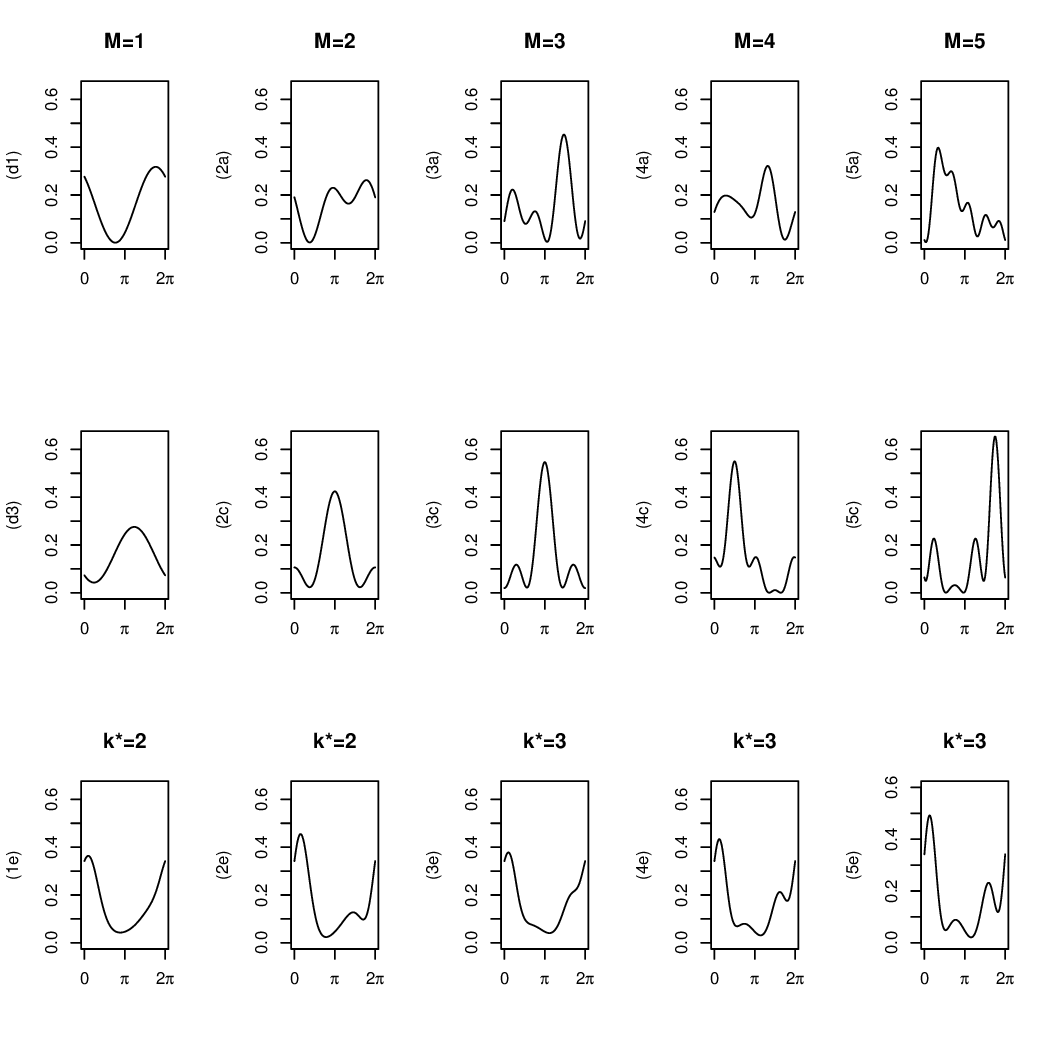}
\caption{Models for simulated data: The first row includes non-symmetric NNTS densities (columns 2 to 5). The second row includes symmetric NNTS densities, ordered by $M$ from 1 to 5. The third row includes k-sine densities: the first two use $k^*=2$ with $\lambda=0.2$ and $\lambda=0.6$, and the last three use $k^*=3$ with $\lambda=0.2, 0.4 \mbox{ and } 0.6$, respectively. 
\label{ressimulaciones}
}
}
\end{figure}

We simulated datasets of varying sample sizes from symmetric and skewed NNTS models to evaluate the rejection rates of the proposed NNTS likelihood ratio ($LR_{GS}$) test of symmetry. 
Additionally, datasets were simulated from the k-sine model with a density function as defined in Equation \ref{ksine}. Figure \ref{ressimulaciones} includes plots of the NNTS and k-sine densities used to generate the simulated datasets. The first two rows show plots of the skewed and symmetric NNTS models, while the third row displays plots of the k-sine densities.

In the first two rows, the columns of Figure \ref{ressimulaciones} are organized from $M=1$ to $M=5$. The first row includes the density functions of skewed (asymmetric) NNTS models, while the second row includes the density functions of symmetric NNTS models in with varying angles of symmetry $\mu$. For the NNTS models in the first two rows, different number of modes are shown. 

The third row of Figure \ref{ressimulaciones} contains plots of the k-sine densities, organized according their parameters $k^*$ and $\lambda$ as defined in Equation \ref{ksine}. The first two plots feature $k^*=2$ with $\lambda=0.2$ and $\lambda=0.6$, respectively. The last three plots represent densities with $k^*=3$ and $\lambda=0.2$, $\lambda=0.4$ and $\lambda=0.6$, respectively.
For the k-sine model, increasing the value of $k^*$ allows for more modes in the density function. Similarly, increasing $\lambda$ produces more skewed density functions.

Table \ref{skewness} displays the values of the sample skewness coefficient, $\hat{s}$, and the proposed NNTS skewness coefficient, $SK_{NNTS}$, for simulated datasets of size 1000 from the
NNTS and k-sine densities in Figure \ref{ressimulaciones}. The NNTS density in plot 3a, for instance, has three modes and a very low $SK_{NNTS}=0.0180$, reflecting minimal skewness, despite a large sample skewness coefficient $\hat{s}=-0.4055$. This is due to a a large concentration of probability mass around the main mode.

For symmetric cases (denoted by c in the second row of Figure \ref{ressimulaciones}), $SK_{NNTS}$ values are below 0.0025. The sample skewness coefficient $\hat{s}$ also shows very small values for these NNTS symmetric cases. For the k-sine densities in the third row of Figure \ref{ressimulaciones}, both $\hat{s}$ and $SK_{NNTS}$ behave as expected, based on the values of the parameters $k^*$ and $\lambda$. Specifically, cases 1e and 3e have small $\hat{s}$ and $SK_{NNTS}$ values, indicating minimal skewness.

Table \ref{simulatedpvalues} presents the p-values of the proposed NNTS likelihood ratio ($LR_{GS}$) reflective symmetry test, in both its asymptotic chi-squared and bootstrapped versions, for the symmetric (c) and skewed (a) NNTS models in Figure \ref{ressimulaciones}. For comparison, it also includes p-values for the bootstrapped version of Pewsey's (2002) $\bar{b}_2$ test, as recommended by Ameijeiras-Alonso et al. (2021) when the mean direction of symmetry, base symmetric model and $k^*$ are unknown. At a significance level of 5\%, for the asymmetric case (a) with $M=2$ (plot $2a$ in Figure \ref{ressimulaciones}), the $\bar{b}_2$ test does not reject the null hypothesis of reflective symmetry even for large sample sizes, such as 1000. However, the asymptotic chi-squared and bootstrapped NNTS likelihood ratio tests only fail to reject the null for a small sample sizes, such as 20. 
For the plot 3a ($M=3$), all three tests ($\bar{b}_2$, $LR_{GS}$ and bootstrapped $LR_{GS}$ BLR) required at least 100 observations to reject the null hypothesis of
reflective symmetry. For cases 4a and 5a, the $LR_{GS}$ and bootstrapped $LR_{GS}$ tests required smaller sample sizes to reject the null hypothesis compared to the $\bar{b}_2$ test.

For the symmetric NNTS cases in the second row of Figure \ref{ressimulaciones}, all three tests produced large p-values. However, for case 2c and small sample sizes of 20 or 50, the $LR_{GS}$ and bootstrapped $LR_{GS}$ erroneously rejected the null hypothesis. This also occured for the $\bar{b}_2$ test with a sample size of 50. Additionally, the $\bar{b}_2$ erroneously rejected the null hypothesis in cases d1 and d2, which correspond to NNTS densities with $M=1$ that are symmetric by definition. Finally, for the skewed k-sine densities in the third row of Figure \ref{ressimulaciones}, all three tests behaved similarly in terms of their p-values. However, for cases 1e and 3e with small skewness, larger sample sizes were required to reject the null hypothesis.

Since the results in Table \ref{simulatedpvalues} are based on only one simulated dataset for each cosidered sample size,
Tables \ref{b2rejectionrates} and \ref{nntsrejectionrates} present rejection rates obtained by simulating 100 datasets for each NNTS and k-sine model in Figure \ref{ressimulaciones}
to analyze the size and power of the $\bar{b}_2$ and $LR_{GS}$ tests. 

In terms of power for the NNTS skewed cases in the first row of Figure \ref{ressimulaciones}, the $LR_{GS}$ is significantly more powerful than the $\bar{b}_2$ test for cases 2a, 4a and 5a, requiring smaller sample sizes to achieve high rejection rates. For case 3a with small skewness, both tests behaved similarly.

To compare the size of both tests, rejection rates for the symmetric NNTS cases in the second row of Figure \ref{ressimulaciones} (denoted as c cases) were analyzed. Both tests showed rejection rates consistent with the significance levels of 10\%, 5\% and 1\%, indicating that the test sizes are adequate. 

Finally, for the skewed k-sine cases in the third row of Figure \ref{ressimulaciones} (denoted as e cases), the $\bar{b}_2$ test exhibited slightly higher power than the $LR_{GS}$ for case 1e
with very small skewness and only two modes. A similar pattern was observed for case 2e with two modes. However, for cases 3e, 4e and 5e which feature three modes, the $LR_{GS}$ demonstrated significantly higher power than the $\bar{b}_2$ test. The difference in power between the tests was considerable, suggesting that the $\bar{b}_2$ test may not be suitable for multimodal datasets. 

To confirm the null asymptotic chi-squared distribution of the $LR_{GS}$ NNTS test, we conducted uniformity tests for the p-values obtained from simulations for each of the symmetric cases (2c to 5c). The uniformity of the p-values was not rejected, confirming the asymptotic chi-squared null distribution of the NNTS test when the sample size is at least $25M$. This rule of thumb can be appled in practical uses of the asymptotic chi-squared $LR_{GS}$ test. For smaller sample sizes (< $25M$), the bootstrapped version of the $LR_{GS}$ test is recommended.   

 \section{Real Examples}

The first example involves to the directions chosen by 100 ants in response to an evenly illuminated black target. This dataset, taken by Fisher (1993, pp. 243) from Jander (1957), is considered to originate from a symmetric circular distribution. The second example includes the directions taken by 76 turtles after treatment. This dataset, obtained by Fisher (1993, pp. 241) from Stephens (1969), is also presumed to originate from a symmetric circular distribution. The third example consists of 214 dragonfly orientations reported by Hisada (1972). Previous analyses have produced controversial results, with some rejecting and other failing to reject the null hypothesis of reflective symmetry. The fourth example, taken from the $R$ package $circular$ (Agostinelli and Lund, 2022), includes 310 wind directions recorded at a monitoring station in the Italian Alps (Col de la Roa). Wind directions were recorded every 15 minutes between 3.00am and 4.00am over 62 consecutive days (January 29 to March 31, 2001). In this case, time autocorrelation between wind directions was not considered and the wind direction-generating process was assumed to be stationary. Previous analyses concluded that the density function of the wind directions is skewed. Parameters for these examples were fitted using the maximum likelihood method implemented in the $R$ package $CircNNTSR$ (Fernández-Durán and Gregorio-Domínguez, 2016).

From the simulation study, we recommend using the value of $M$ that corresponds to the best Bayesian Information Criterion (BIC) symmetric NNTS model for the practical application of the NNTS likelihood ratio test for reflective symmetry.

\subsection{Ant data}

\begin{figure}[h]
\center{
\includegraphics[scale=.5, bb= 0 0 504 504]{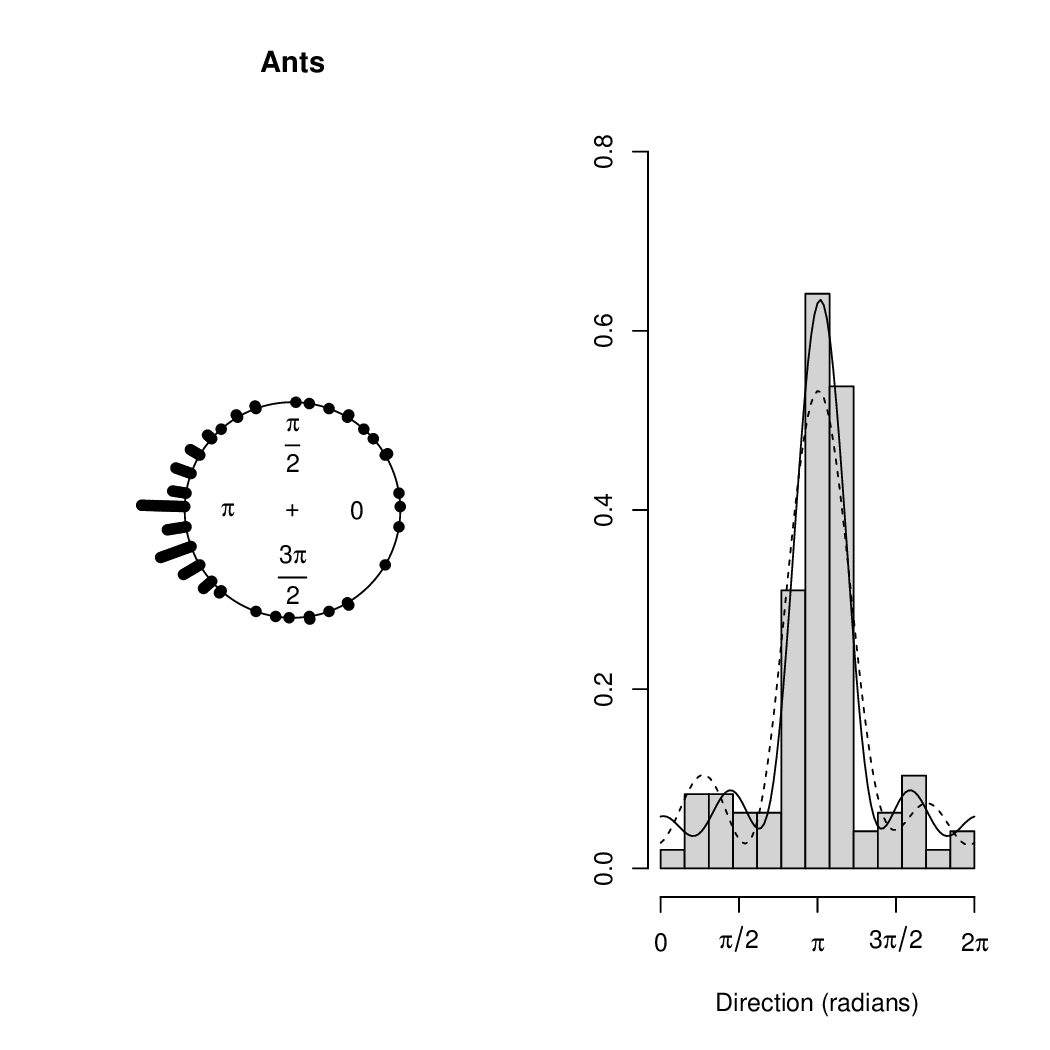}
\caption{Ant data (100 observations). Left plot: Directions chosen by 100 ants in response to an evenly illuminated black target (Fisher, 1993). 
Right plot: Histogram and best BIC-fitted NNTS densities: general ($M=3$, dashed line) and symmetric ($M=4$, solid line).
\label{Rawants}
}
}
\end{figure}

Figure \ref{Rawants} displays the raw data plot of the reduced ant dataset, consisting of 100 observations.
This plot reveals a clear directional preference at approximately $180^o$. Table \ref{antsloglikAICBIC} provides the log-likelihood values, 
Akaike's Information Criterion (AIC), and Bayesian Information Criterion (BIC) for both general and 
symmetric NNTS models, with $M=0, 1, \ldots, 5$. Additionally, the observed values of the likelihood
ratio test statistic for symmetry ($LR_{GS}$) and the corresponding asymptotic chi-squared and bootstrapped p-values are included.

The optimal general and symmetric NNTS models, based on BIC, are those with $M=3$ and $M=4$, respectively. Applying the likelihood ratio test for symmetry for $M=4$ yields p-values of 0.585 and 0.648, for the asymptotic chi-squared and bootstrapped NNTS likelihood ratio tests, respectively. These results suggest that the null hypothesis of symmetry cannot be rejected for either case. 

Figure \ref{Rawants} illustrates the best BIC general ($M=3$) and symmetric ($M=4$) NNTS densities for the ant data.
The bootstrapped version of the $\bar{b}_2$ test also supports the null hypothesis of reflective symmetry, with a p-value of 0.493.

\begin{figure}[h]
\center{
\includegraphics[scale=.5, bb= 0 0 504 504]{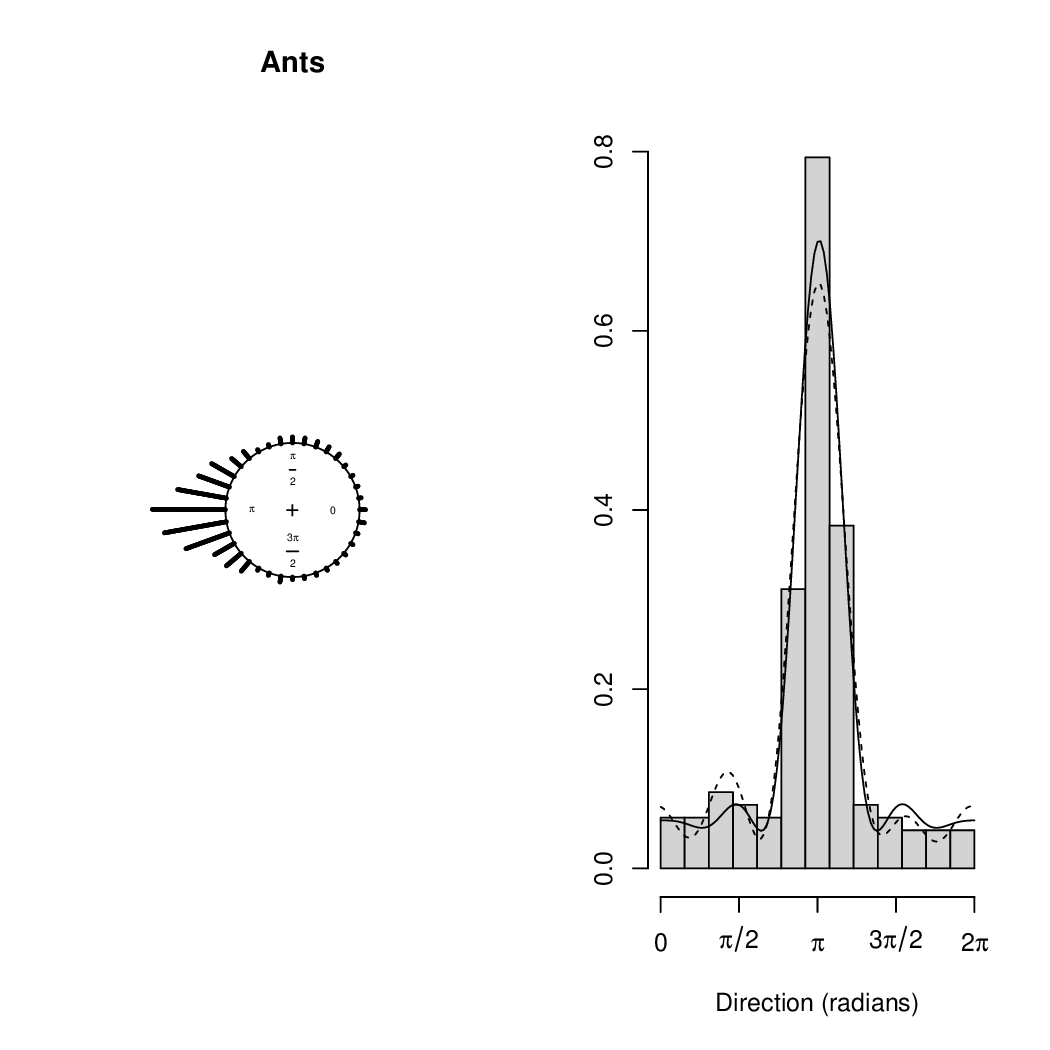}
\caption{Ant data (730 observations). Left plot: Directions chosen by 730 ants in response to an evenly illuminated black target. 
Right plot: Histogram and BIC-fitted NNTS densities: general ($M=4$, dashed line) and symmetric ($M=5$, solid line).
\label{Rawants730}
}
}
\end{figure}

For further analysis, we expanded the dataset to include 730 observations, as illustrated in Figure \ref{Rawants730}. This enlarged dataset was previously analysed by Pewsey (2004) and Ley and Verdebout (2014), who considered a known angle of symmetry equal to $180^o$. Pewsey (2004) obtained a p-value of 0.011 using the $b_2^*$ test, and Ley and Verdebout (2014) rejected the null hypothesis of reflective symmetry. Table \ref{antsloglikAICBIC730} summarizes the results of the NNTS tests for $M=0,1, \ldots, 10$, detailing the corresponding log-likelihoods, AIC, and BIC values. The best BIC NNTS symmetric model is identified with $M=5$, yielding p-values of 
0.059 and 0.069 for the asymptotic chi-squared and bootstrapped NNTS likelihood ratio tests, respectively.These results do not reject the null hypothesis of reflective symmetry at a 5\% significance level. Additionally, the bootstrapped $\bar{b}_2$ test produces a p-value of 0.018.

\subsection{Turtle data}

\begin{figure}[h]
\center{
\includegraphics[scale=.5, bb= 0 0 504 504]{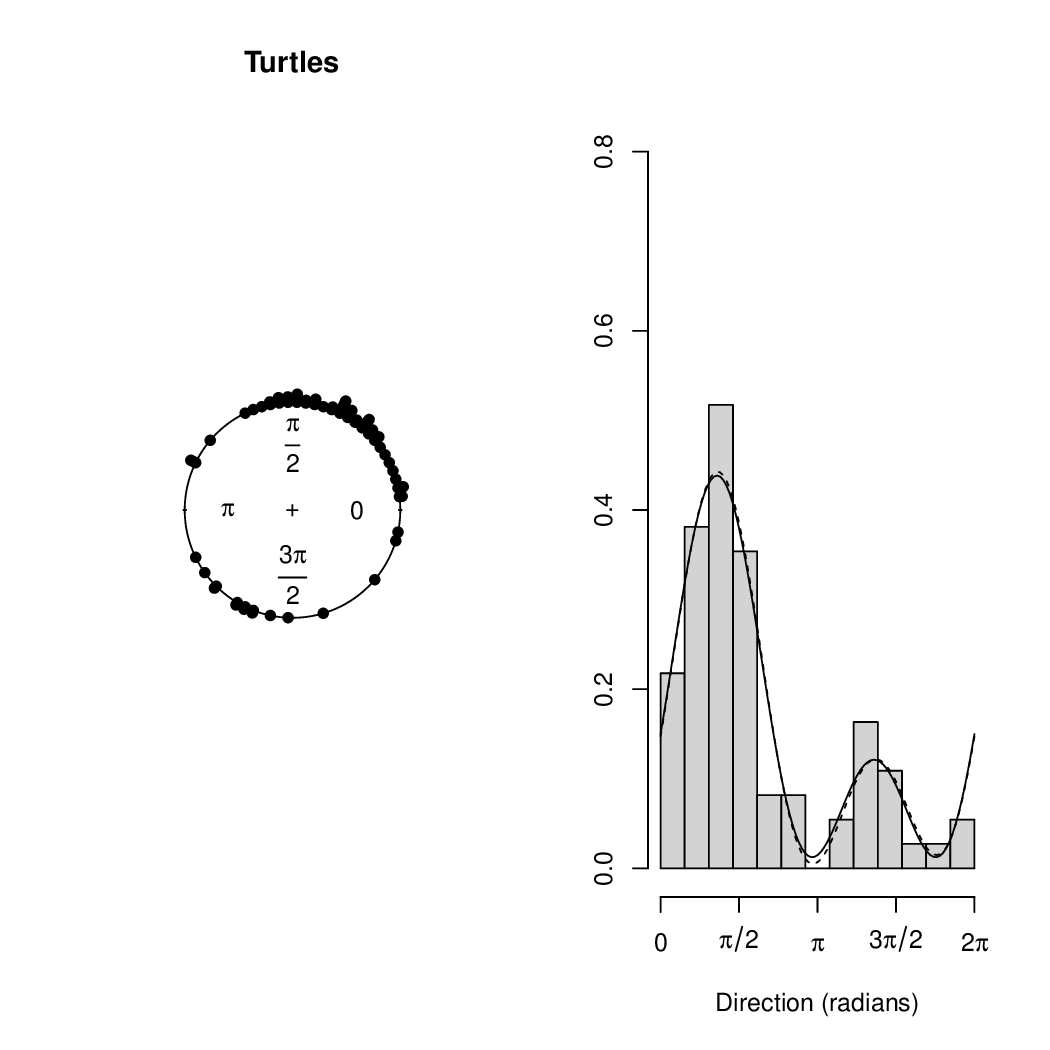}
\caption{Turtle data. Left plot: Directions taken by 76 turtles after treatment (Stephens, 1969). Right plot: Histogram and the best BIC fitted general (dashed line) and symmetric (solid line) NNTS densities, both with $M=2$.
\label{Rawturtles}
}
}
\end{figure}

Figure \ref{Rawturtles} displays the raw data plot, which, as noted by Fisher (1993 pp. 23), suggests a degree of bimodality. 
Table \ref{turtlesloglikAICBIC} summarizes the log-likelihood values, Akaike's Information Criterion (AIC), and Bayesian Information Criterion (BIC) for both general and symmetric NNTS models. 
Additionally, the table includes the likelihood ratio test statistic ($LR_{GS}$) and the corresponding asymptotic chi-squared and bootstrapped p-values for $M=0, \ldots, 4$. 

The results in Table \ref{turtlesloglikAICBIC} indicate no evidence
against the hypothesis of reflective symmetry hypothesis. Moreover, Figure \ref{Rawturtles} shows that the best BIC-fitted general ($M=2$) and symmetric ($M=2$) NNTS densities are nearly
identical, corroborating the results of the likelihood ratio test for reflective symmetry. Similarly, the bootstrapped version of the $\bar{b}_2$ test fails to reject the null hypothesis
of reflective symmetry, yielding a p-value equal of 0.764.

\subsection{Dragonfly data}

\begin{figure}[h]
\center{
\includegraphics[scale=.5, bb= 0 0 504 504]{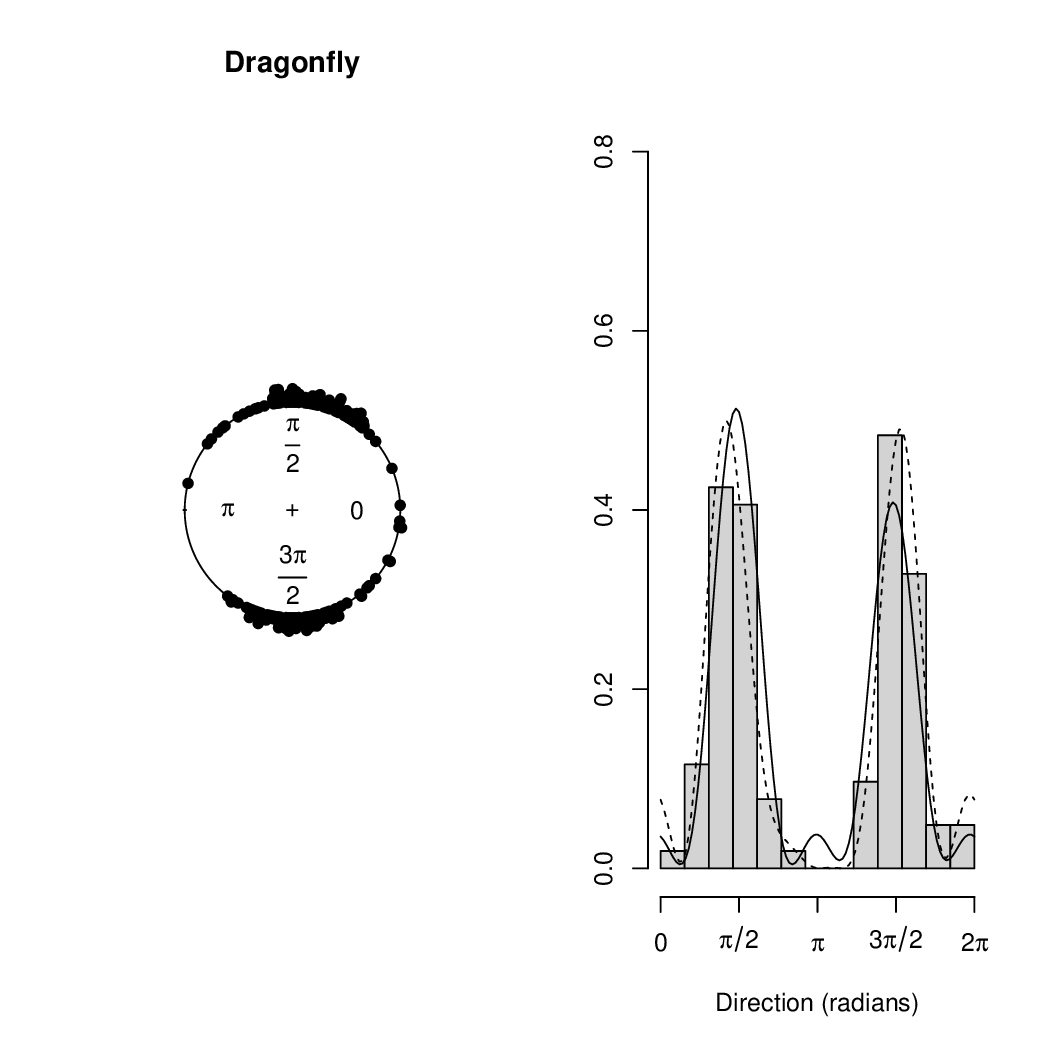}
\caption{Dragonfly data. Left plot: Orientations of 214 dragonflies, as reported by Hisada (1972). Right plot: Histogram and best BIC-fitted densities: general with ($M=5$, dashed line) and symmetric ($M=4$, solid line).
\label{Rawdragonfly}
}
}
\end{figure}

The dataset consists of 214 dragonfly orientations relative to the Sun's azimuth, as reported by Hisada (1972) and analyzed by Pewsey (2004).
Figure \ref{Rawdragonfly} displays the histogram and density plots for the best BIC general ($M=5$) and symmetric ($M=4$) NNTS models.
 
Upon dividing the observations into two intervals (0 to $\pi$ and $\pi$ to 2$\pi$), some skewness is observed in the histogram.
The bootstrapped $\bar{b}_2$ test yields a high p-value of 0.694. Previously, Pewsey (2004) applied the $b_2^*$ test to this dataset, considering 
a known symmetry angle of 0, and reported a p-value of 0.020. Using the NNTS likelihood ratio test  with $M=4$, corresponding to the best BIC symmetric
NNTS model, we obtain an asymptotic chi-squared p-value of 0.025 and a bootstrapped p-value of 0.084. For the dragonfly dataset, the conclusions derived from the
$\bar{b}_2$, $b_2^*$ and NNTS likelihood ratio test differ significantly. Considering the p-values of the NNTS likelihood ratio tests for $M \ne 4$, the null hypothesis of reflective symmetry is also rejected at the 5\% significance level. These findings highlights the limitations of the $\bar{b}_2$ test for multimodal datasets
and emphasize the advantages of the proposed NNTS test.

\subsection{Wind directions data}

\begin{figure}[h]
\center{
\includegraphics[scale=.5, bb= 0 0 504 504]{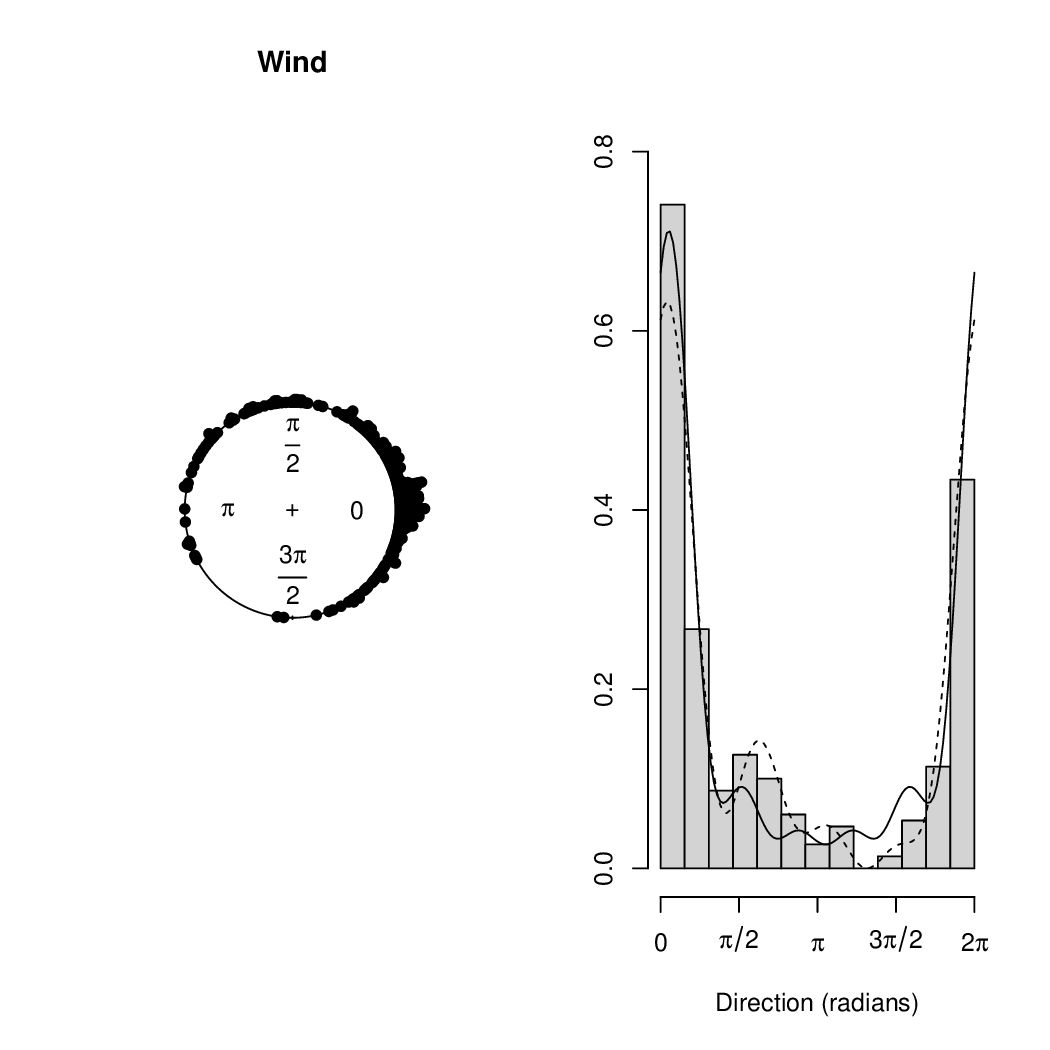}
\caption{Wind data. Left plot: Wind directions recorded at a monitoring station in the Italian Alps (Col de la Roa). Right plot: Histogram and best BIC-fitted NNTS densities: general ($M=4$, dashed line) and symmetric ($M=5$, solid line).
\label{Rawwind}
}
}
\end{figure}

Figure \ref{Rawwind} presents the raw data plot, which shows a skewed histogram. Table \ref{windloglikAICBIC} provides p-values below 0.001 for all tests conducted with values of $M$ ranging from
2 to 12, including the best BIC general NNTS model ($M=4$) and the best BIC symmetric model ($M=5$). The bootstrapped $\bar{b}_2$ with 999 simulations yields a p-value of 0.001,
rejecting the null hypothesis of reflective symmetry.

\subsection{Time of gun crimes data}

\begin{figure}[h]
\center{
\includegraphics[scale=.5, bb= 0 0 504 504]{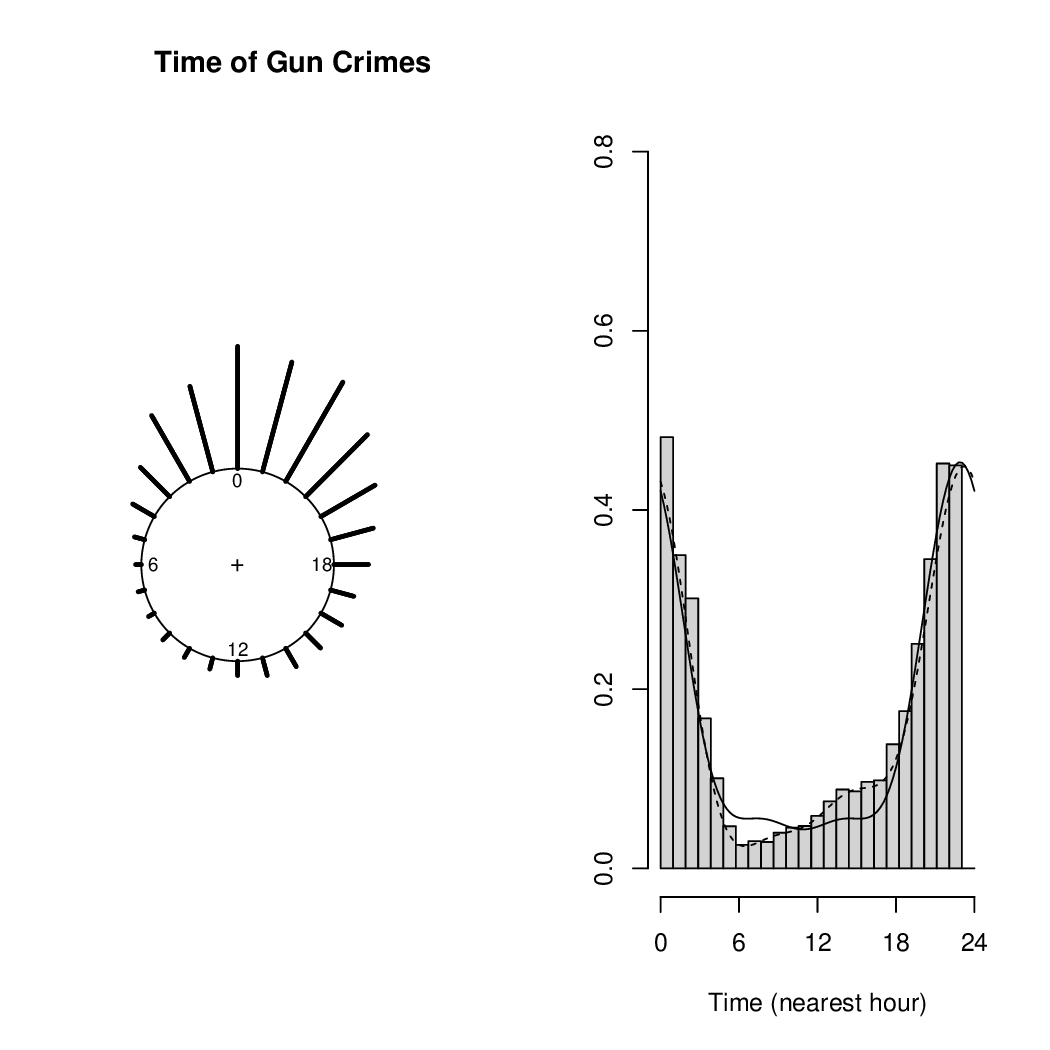}
\caption{Time of gun crimes data. Left plot: Times (nearest hour) of gun crimes in Pittsburgh, PA, from January 1992 to May 1996. Each point represents 15 observations. Right plot: Histogram and best BIC-fitted NNTS densities: general ($M=3$, dashed line) and symmetric ($M=3$, solid line).
\label{guncrimes}
}
}
\end{figure}

Cohen and Gorr (2006) reported data on the times of gun crimes in Pittsburgh, PA, recorded from January 1992 to May 1996. The times were noted to the nearest hour, with a total of 15,831 observations. Figure \ref{guncrimes} includes the circular dot plot and histogram of these observations, where a high degree of skewness is evident. Both the general and symmetric NNTS models use $M=3$, and the null hypothesis of reflective symmetry is strongly rejected. The asymptotic chi-squared p-value of 3.09e-120 and bootstrapped NNTS and $\bar{b}_2$ test p-values are both 0.001 with 999 simulations.

\section{Conclusions}

The proposed NNTS likelihood ratio reflective symmetry test demonstrates an excellent performance in detecting symmetry in circular distributions, even for multimodal datasets. This is evident
form its application to real datasets on ant and turtle movements, dragonfly orientations, wind directions and, times of gun crimes. The proposed tests are robust regarding the selection of the parameter $M$, ensuring reliable practical implementation. When testing for reflective symmetry in circular data, it is recommended to apply the NNTS likelihood ratio test for reflective symmetry when using NNTS models. This approach allows consideration of NNTS symmetric models, which are more parsimonious than general NNTS models, as the number of parameters is reduced from $2M$ for general
NNTS models to $M+1$ for NNTS symmetric models (for $M>2$).

From the simulation study, comparing the proposed NNTS likelihood ratio test with the $\bar{b}_2$ test (recommended for cases with an unknown angle of symmetry) reveals that the NNTS test has higher power, particularly for asymmetric multimodal circular data. Additionally, for NNTS models with $M=1$, which are symmetric by definition, the bootstrapped version of the $\bar{b}_2$ test can reject the null hypothesis of reflective symmetry even with large sample sizes. For the practical application of the NNTS likelihood ratio test for reflective symmetry, it is advised to use the value of $M$ that corresponds to the best BIC NNTS symmetric model.

\newpage

\renewcommand{\baselinestretch}{1}

\begin{table}[t]
\begin{center}
\scalebox{0.6}{
\begin{tabular}{llllcc}
\hline
$M$     & $k^*$  & $\lambda$ & Plot & $\hat{s}$ & $SK_{NNTS}$ \\
\hline
\multicolumn{6}{l}{(a) Non-symmetric case. Plots of NNTS densities in the first row of Figure \ref{ressimulaciones} (columns 2 to 5).} \\
 \hline
2 & & &  2a & -0.0265  & 0.1957  \\
3 & & &  3a & -0.4055  & 0.0180  \\
4 & & &  4a &  0.3432  & 0.1096  \\
5 & & &  5a & -0.1921  & 0.1789  \\
\hline
\multicolumn{6}{l}{(c) Symmetric case. Plots of NNTS densities in the second row of Figure \ref{ressimulaciones} (columns 2 to 5).} \\
\hline
2 & & &  2c &  0.0225  & 0.0008  \\
3 & & &  3c &  0.0390  & 0.0021  \\
4 & & &  4c & -0.0017  & 0.0005  \\
5 & & &  5c &  0.0638  & 0.0017  \\
\hline
\multicolumn{6}{l}{(d) $M=1$ (Symmetric by definition). Plots of NNTS densities in the first two rows of the first column of Figure \ref{ressimulaciones}.} \\
\hline
1 & & &  d1 &  0.0984  & 0  \\
1 & & &  d3 & -0.0677  & 0  \\
\hline
\multicolumn{6}{l}{(e) Non-symmetric case. Plots of $k^*$-sine von Mises ($\mu=0$ and $\kappa=1$) densities in the third row of Figure \ref{ressimulaciones}.} \\
\hline
 & 2 & 0.2 & 1e &   0.1453  & 0.0052 \\
 & 2 & 0.6 & 2e &   0.2662  & 0.0242 \\
 & 3 & 0.2 & 3e &  -0.1005  & 0.0047 \\
 & 3 & 0.4 & 4e &  -0.1914  & 0.0247 \\
 & 3 & 0.6 & 5e &  -0.3268  & 0.0471 \\
\hline
\end{tabular}}
\caption{Simulated data: Sample skewness coefficient ($\hat{s}$) and the proposed NNTS skewness coefficient ($SK_{NNTS}$) for simulated dataset of size 1000, from the
NNTS densities shown in Figure \ref{ressimulaciones}. \label{skewness} }
\end{center}
\end{table}

\begin{table}[t]
\begin{center}
\scalebox{0.5}{
\begin{tabular}{lllc||cccccc||cccccc||cccccc}
\hline
        &       & & & \multicolumn{6}{|c||}{$\bar{b}_2$ Bootstrapped p-value} & \multicolumn{6}{c||}{NNTS LR $\chi^2$ Asymptotic p-value} & \multicolumn{6}{c}{NNTS BLR p-value}  \\
$M$     &$k^*$  &$\lambda$&Plot& 20  & 50  & 100  & 200  & 500  & 1000  & 20  & 50  & 100  & 200  & 500  & 1000  & 20  & 50  & 100  & 200  & 500  & 1000  \\
\hline
\multicolumn{22}{l}{(a) Non-symmetric case. Plots of NNTS densities in the first row of Figure \ref{ressimulaciones} (columns 2 to 5).} \\
 \hline
2 &  &  & 2a & 0.123 & 0.022 & 0.930 & 0.659 & 0.322 & 0.793 & 0.411 & <0.001 & <0.001 & <0.001 & <0.001 & <0.001 & 0.109 & 0.001 & 0.001 & 0.001 & 0.001 & 0.001 \\
3 &  &  & 3a & 0.533 & 0.275 & 0.050 & 0.002 & 0.002 & 0.001 & 0.624 & 0.409  & 0.028  & 0.007  & <0.001 & <0.001 & 0.410 & 0.265 & 0.019 & 0.005 & 0.001 & 0.001 \\
4 &  &  & 4a & 0.045 & 0.004 & 0.336 & 0.085 & 0.013 & 0.001 & 0.130 & 0.195  & 0.293  & 0.005  & <0.001 & <0.001 & 0.122 & 0.371 & 0.466 & 0.024 & 0.001 & 0.001 \\
5 &  &  & 5a & 0.008 & 0.630 & 0.976 & 0.234 & 0.055 & 0.001 & 0.034 & <0.001 & <0.001 & <0.001 & <0.001 & <0.001 & 0.037 & 0.001 & 0.006 & 0.001 & 0.001 & 0.001 \\
\hline
\multicolumn{22}{l}{(c) Symmetric case. Plots of NNTS densities in the second row of Figure \ref{ressimulaciones} (columns 2 to 5).} \\
\hline
2 &  &  & 2c & 0.741 & 0.018 & 0.071 & 0.234 & 0.366 & 0.071 & 0.019 & 0.023 & 0.066 & 0.294 & 0.501 & 0.769 & 0.021 & 0.008 & 0.073 & 0.317 & 0.520 & 0.789 \\
3 &  &  & 3c & 0.146 & 0.414 & 0.477 & 0.963 & 0.497 & 0.651 & 0.169 & 0.668 & 0.195 & 0.514 & 0.926 & 0.383 & 0.170 & 0.471 & 0.159 & 0.495 & 0.935 & 0.395 \\
4 &  &  & 4c & 0.668 & 0.604 & 0.402 & 0.619 & 0.402 & 0.983 & 0.844 & 0.816 & 0.651 & 0.992 & 0.830 & 0.839 & 0.531 & 0.631 & 0.521 & 0.976 & 0.727 & 0.719 \\
5 &  &  & 5c & 0.246 & 0.302 & 0.167 & 0.196 & 0.765 & 0.271 & 0.346 & 0.720 & 0.723 & 0.620 & 0.657 & 0.776 & 0.150 & 0.450 & 0.568 & 0.612 & 0.583 & 0.704 \\
\hline
\multicolumn{22}{l}{(d) $M=1$ (Symmetric by definition). Plots of NNTS densities in the first two rows of the first column of Figure \ref{ressimulaciones}.} \\
\hline
1 & &&d1& 0.868 & 0.490 & 0.346 & 0.774 & 0.023 & 0.120 &          &       &       &       &       &       &       &       &          &       &       &       \\
1 & &&d3& 0.686 & 0.017 & 0.848 & 0.437 & 0.035 & 0.120 &          &       &       &       &       &       &       &       &          &       &       &       \\
\hline
\multicolumn{22}{l}{Power: (e) Non-symmetric case. Plots of $k^*$-sine von Mises ($\mu=0$ and $\kappa=1$) densities in the third row of Figure \ref{ressimulaciones}.} \\
\hline
 & 2 & 0.2 & 1e & 0.187 & 0.713 & 0.808 & 0.126 & 0.006 & 0.006 & 0.073 & 0.757 &  0.175 & 0.040  &  0.004 &  0.001 & 0.033 & 0.813 & 0.189 & 0.057 & 0.007 & 0.001 \\
 & 2 & 0.6 & 2e & 0.794 & 0.082 & 0.001 & 0.001 & 0.001 & 0.001 & 0.178 & 0.109 & <0.001 & <0.001 & <0.001 & <0.001 & 0.193 & 0.111 & 0.001 & 0.001 & 0.001 & 0.001 \\
 & 3 & 0.2 & 3e & 0.147 & 0.659 & 0.933 & 0.330 & 0.356 & 0.187 & 0.061 & 0.900 &  0.665 & 0.618  & 0.050  & 0.009  & 0.046 & 0.921 & 0.728 & 0.647 & 0.068 & 0.008 \\
 & 3 & 0.4 & 4e & 0.841 & 0.278 & 0.179 & 0.006 & 0.024 & 0.003 & 0.132 & 0.290 &  0.144 & 0.001  & <0.001 & <0.001 & 0.148 & 0.308 & 0.233 & 0.001 & 0.001 & 0.001 \\
 & 3 & 0.6 & 5e & 0.648 & 0.320 & 0.027 & 0.001 & 0.001 & 0.001 & 0.103 & 0.010 & <0.001 & <0.001 & <0.001 & <0.001 & 0.052 & 0.120 & 0.001 & 0.001 & 0.001 & 0.001 \\
\hline
\end{tabular}}
\caption{Simulated data: P-values of the bootstrapped $\bar{b}_2$ test with 999 bootstrap simulations and the NNTS likelihood ratio test, including asymptotic chi-squared and bootstrapped (with 999 simulations) p-values, for reflective symmetry. \label{simulatedpvalues} }
\end{center}
\end{table}

\begin{table}[t]
\begin{center}
\scalebox{0.6}{
\begin{tabular}{lllc||cccccc||cccccc||cccccc}
\hline
   &      && & \multicolumn{6}{|c||}{$\alpha$=10\%} & \multicolumn{6}{c||}{$\alpha$=5\%} & \multicolumn{6}{c}{$\alpha$=1\%}  \\
$M$&$k^*$ &$\lambda$& Plot& 20  & 50  & 100  & 200  & 500  & 1000  & 20  & 50  & 100  & 200  & 500  & 1000  & 20  & 50  & 100  & 200  & 500  & 1000  \\
\hline
\multicolumn{22}{l}{Power: (a) Non-symmetric case. Plots of NNTS densities in the first row of Figure \ref{ressimulaciones} (columns 2 to 5).} \\
 \hline
2 &  &  & 2a & 16 & 16 & 21 & 22 & 35 &  68 & 10 & 12 & 13 & 15 & 23 &  54 & 4 &  8 &  8 & 10 & 14 &  29 \\
3 &  &  & 3a & 19 & 35 & 50 & 68 & 96 & 100 & 12 & 32 & 40 & 59 & 90 & 100 & 4 & 22 & 32 & 41 & 84 & 100 \\
4 &  &  & 4a & 21 & 22 & 34 & 38 & 81 &  97 & 15 & 16 & 28 & 35 & 73 &  92 & 8 &  6 & 19 & 31 & 60 &  85 \\
5 &  &  & 5a &  8 & 18 & 39 & 56 & 87 &  97 &  4 & 11 & 28 & 42 & 82 &  96 & 1 &  2 & 13 & 20 & 59 &  89 \\
\hline
\multicolumn{22}{l}{Level: (c) Symmetric case. Plots of NNTS densities in the second row of Figure \ref{ressimulaciones} (columns 2 to 5).} \\
\hline
2 &  &  & 2c &  3 & 14 & 10 & 10 &  9 & 15 &  2 & 5 & 6 & 8 & 4 & 7 & 0 & 2 & 0 & 2 & 1 & 1 \\
3 &  &  & 3c &  6 & 11 & 10 &  8 & 11 & 11 & 12 & 6 & 5 & 5 & 6 & 3 & 4 & 4 & 1 & 1 & 0 & 0 \\
4 &  &  & 4c &  8 & 10 & 15 &  7 &  9 &  8 &  3 & 5 & 9 & 4 & 6 & 1 & 0 & 1 & 0 & 0 & 3 & 1 \\
5 &  &  & 5c & 10 & 13 & 10 &  8 & 12 & 11 &  8 & 8 & 5 & 2 & 7 & 8 & 1 & 2 & 2 & 1 & 0 & 2 \\
\hline
\multicolumn{22}{l}{Power: (e) Non-symmetric case. Plots of $k^*$-sine von Mises ($\mu=0$ and $\kappa=1$) densities in the third row of Figure \ref{ressimulaciones}.} \\
\hline
 & 2 & 0.2 & 1e &  8 & 17 &  32 &  50 &  85 &  98 & 6 &  9 & 20 &  41 &  75 &  98 & 1 & 3 &  4 & 20 &  48 &  87 \\
 & 2 & 0.6 & 2e &  0 &  0 & 100 & 100 & 100 & 100 & 0 &  0 & 99 & 100 & 100 & 100 & 0 & 0 &  0 & 99 & 100 & 100 \\
 & 3 & 0.2 & 3e &  6 & 15 &  15 &  28 &  40 &  70 & 4 & 11 &  6 &  16 &  29 &  61 & 1 & 0 &  1 &  7 &  14 &  32 \\
 & 3 & 0.4 & 4e & 10 & 24 &  35 &  47 &  90 &  99 & 9 & 14 & 23 &  38 &  76 &  97 & 5 & 5 &  9 & 25 &  51 &  94 \\
 & 3 & 0.6 & 5e & 12 & 32 &  49 &  73 & 100 & 100 & 5 & 23 & 36 &  62 &  99 & 100 & 1 & 7 & 18 & 38 &  92 & 100 \\
\hline
\end{tabular}}
\caption{Bootstrapped $\bar{b}_2$ Test: Rejection rates to analyze the power and size of the test. \label{b2rejectionrates} }
\end{center}
\end{table}

\begin{table}[t]
\begin{center}
\scalebox{0.6}{
\begin{tabular}{lllc||cccccc||cccccc||cccccc}
\hline
    &       &          &     & \multicolumn{6}{|c||}{$\alpha$=10\%} & \multicolumn{6}{c||}{$\alpha$=5\%} & \multicolumn{6}{c}{$\alpha$=1\%}  \\
$M$ & $k^*$ & $\lambda$& Plot& 20  & 50  & 100  & 200  & 500  & 1000  & 20  & 50  & 100  & 200  & 500  & 1000  & 20  & 50  & 100  & 200  & 500  & 1000  \\
\hline
\multicolumn{22}{l}{Power: (a) Non-symmetric case. Plots of NNTS densities in the first row of Figure \ref{ressimulaciones} (columns 2 to 5).} \\
 \hline
2 &  &  & 2a & 40 & 93 & 100 & 100 & 100 & 100 & 28 & 89 & 100 & 100 & 100 & 100 &  5 & 67 & 97 & 100 & 100 & 100 \\
3 &  &  & 3a & 27 & 38 &  51 &  68 &  95 & 100 & 18 & 28 &  44 &  58 &  93 & 100 &  9 & 16 & 23 &  41 &  85 &  99 \\
4 &  &  & 4a & 41 & 69 &  81 &  98 & 100 & 100 & 30 & 59 &  75 &  95 & 100 & 100 & 11 & 40 & 56 &  87 &  99 & 100 \\
5 &  &  & 5a & 39 & 91 &  98 & 100 & 100 & 100 & 24 & 83 &  97 & 100 & 100 & 100 & 14 & 66 & 94 &  99 & 100 & 100 \\
\hline
\multicolumn{22}{l}{Level: (c) Symmetric case. Plots of NNTS densities in the second row of Figure \ref{ressimulaciones} (columns 2 to 5).} \\
\hline
2 &  &  & 2c &  4 &  8 & 14 &  9 & 10 & 12 &  0 &  3 & 6 & 7 & 4 & 5 &  0 & 1 & 0 & 1 & 0 & 0 \\
3 &  &  & 3c &  3 & 14 & 10 & 16 & 13 &  8 &  1 &  9 & 5 & 7 & 6 & 4 &  1 & 2 & 1 & 1 & 1 & 1 \\
4 &  &  & 4c & 15 & 13 &  6 &  7 &  5 &  5 &  9 &  7 & 4 & 3 & 3 & 3 &  4 & 1 & 0 & 2 & 1 & 0 \\
5 &  &  & 5c & 19 & 16 & 11 &  6 &  8 &  9 & 14 & 11 & 8 & 4 & 6 & 3 & 10 & 4 & 1 & 1 & 2 & 0 \\
\hline
\multicolumn{22}{l}{Power: (e) Non-symmetric case. Plots of $k^*$-sine von Mises ($\mu=0$ and $\kappa=1$) densities in the third row of Figure \ref{ressimulaciones}.} \\
\hline
 & 2 & 0.2 & 1e & 11 & 27 & 34 & 50 &  76 &  98 &  7 & 18 & 23 & 36 &  66 &  95 &  2 &  5 & 11 & 18 &  43 &  85 \\
 & 2 & 0.6 & 2e & 15 & 46 & 85 & 98 & 100 & 100 &  7 & 38 & 76 & 97 & 100 & 100 &  0 & 16 & 57 & 90 & 100 & 100 \\
 & 3 & 0.2 & 3e & 25 & 40 & 48 & 64 &  87 &  99 & 14 & 32 & 37 & 57 &  82 &  97 &  4 & 16 & 26 & 32 &  65 &  92 \\
 & 3 & 0.4 & 4e & 25 & 54 & 72 & 92 & 100 & 100 & 18 & 44 & 57 & 88 & 100 & 100 &  8 & 23 & 40 & 73 & 100 & 100 \\
 & 3 & 0.6 & 5e & 32 & 56 & 85 & 95 & 100 & 100 & 22 & 44 & 76 & 95 & 100 & 100 & 10 & 25 & 54 & 80 & 100 & 100 \\
\hline
\end{tabular}}
\caption{NNTS asymptotic chi-squared likelihood ratio test: Rejection rates to evaluate the power and size of the test. \label{nntsrejectionrates} }
\end{center}
\end{table}

\begin{table}[t]
\begin{center}
\scalebox{0.6}{
\begin{tabular}{c||ccc||ccc||c||ccc||cccc}
\hline
     & \multicolumn{3}{|c||}{General NNTS model} & \multicolumn{3}{c||}{Symmetric NNTS model} &              & \multicolumn{3}{c||}{Likelihood Ratio}     &    \\
$M$  & loglik & AIC & BIC                        & loglik & AIC & BIC                         & $\hat{\mu}$  & $LR_{GS}$ & $\chi^2$ p-value & BLR p-value & $SK_{NNTS}$    \\
\hline
\hline
0 & -183.79 & 367.58  & 367.58 &         &        &        &       &       &       &       &       \\
1 & -153.65 & 311.29  & 316.50 & -153.65 & 311.29 & 316.50 & 3.091 &       &       &       &       \\ 
2 & -141.66 & 291.31  & 301.73 & -141.96 & 289.91 & 297.73 & 3.299 & 0.600 & 0.439 & 0.479 & 0.0037 \\ 
3 & -133.42 & 278.83  & 294.46*& -133.76 & 275.53 & 285.95 & 3.160 & 0.696 & 0.706 & 0.747 & 0.0045 \\ 
4 & -129.32 & 274.64  & 295.48 & -130.29 & 270.58 & 283.60*& 3.198 & 1.937 & 0.585 & 0.648 & 0.0095 \\ 
5 & -126.81 & 273.62  & 299.67 & -129.73 & 271.45 & 287.08 & 3.248 & 5.831 & 0.212 & 0.259 & 0.0604 \\ 
\hline
\end{tabular}}
\caption{Ant data (100 observations): Values of the NNTS parameter ($M$), log-likelihood, AIC, and BIC for general (columns 2-4) and symmetric NNTS (columns 5-7) models.
Estimated angle of reflection ($\hat{\mu}$) is in column 8, loglikelihood ratio test statistic ($LR_{GS}$) in column 9, with its asymptotic chi-squared and bootstrapped ($BLR$)
p-values in columns 10 and 11. Column 12 contains the Wald-type symmetry measure ($SK_{NNTS}$).
\label{antsloglikAICBIC}}
\end{center}
\end{table}

\begin{table}[t]
\begin{center}
\scalebox{0.6}{
\begin{tabular}{c||ccc||ccc||c||ccc||cccc}
\hline
     & \multicolumn{3}{|c||}{General NNTS model} & \multicolumn{3}{c||}{Symmetric NNTS model} &              & \multicolumn{3}{c||}{Likelihood Ratio}     &    \\
$M$  & loglik & AIC & BIC                        & loglik & AIC & BIC                         & $\hat{\mu}$  & $LR_{GS}$ & $\chi^2$ p-value & BLR p-value & $SK_{NNTS}$    \\
\hline
\hline
0 & -1341.65 & 2683.30 & 2683.30 &         &         &         &          &           &           &          &          \\
1 & -1122.85 & 2249.70 & 2258.89 &-1122.85 & 2249.70 & 2258.89 &    3.072 &           &           &          &          \\
2 & -1015.56 & 2039.12 & 2057.50 &-1018.84 & 2043.68 & 2057.46 &    3.148 &    6.561  &    0.010  &   0.013  &   0.0046 \\
3 &  -957.52 & 1927.03 & 1954.59 & -959.96 & 1927.91 & 1946.29 &    3.211 &    4.897  &    0.086  &   0.083  &   0.0038 \\
4 &  -919.45 & 1854.89 & 1891.64*& -923.71 & 1857.41 & 1880.38 &    3.167 &    8.544  &    0.036  &   0.032  &   0.0056 \\
5 &  -915.32 & 1850.64 & 1896.57 & -919.86 & 1851.71 & 1879.27*&    3.173 &    9.082  &    0.059  &   0.069  &   0.0059 \\
6 &  -913.00 & 1850.00 & 1905.11 & -917.65 & 1849.30 & 1881.45 &    3.179 &    9.330  &    0.097  &   0.128  &   0.0062 \\
7 &  -912.69 & 1853.38 & 1917.68 & -918.01 & 1852.03 & 1888.77 &    3.179 &   10.662  &    0.099  &   0.190  &   0.0071 \\
8 &  -910.29 & 1852.58 & 1926.07 & -915.45 & 1848.91 & 1890.24 &    3.173 &   10.322  &    0.171  &   0.266  &   0.0068 \\
9 &  -907.42 & 1850.84 & 1933.52 & -912.50 & 1845.00 & 1890.93 &    3.173 &   10.156  &    0.254  &   0.381  &   0.0067 \\
10&  -906.07 & 1852.14 & 1944.00 & -911.36 & 1844.73 & 1895.25 &    3.173 &   10.583  &    0.305  &   0.464  &   0.0070 \\
\hline
\end{tabular}}
\caption{Ant data (730 observations): Values of the NNTS parameter ($M$), log-likelihood, AIC, and BIC for general (columns 2-4) and symmetric NNTS (columns 5-7) models.
Estimated angle of reflection ($\hat{\mu}$) is in column 8, loglikelihood ratio test statistic ($LR_{GS}$) in column 9, with its asymptotic chi-squared and bootstrapped ($BLR$)
p-values in columns 10 and 11. Column 12 contains the Wald-type symmetry measure ($SK_{NNTS}$).
\label{antsloglikAICBIC730}}
\end{center}
\end{table}

\begin{table}[t]
\begin{center}
\scalebox{0.6}{
\begin{tabular}{c||ccc||ccc||c||ccc||cccc}
\hline
     & \multicolumn{3}{|c||}{General NNTS model} & \multicolumn{3}{c||}{Symmetric NNTS model} &              & \multicolumn{3}{c||}{Likelihood Ratio}     &    \\
$M$  & loglik & AIC & BIC                        & loglik & AIC & BIC                         & $\hat{\mu}$  & $LR_{GS}$ & $\chi^2$ p-value & BLR p-value & $SK_{NNTS}$ \\
\hline
\hline
0 & -139.68 & 279.36 & 279.36 &         &        &        &       &       &       &       &         \\
1 & -126.33 & 256.66 & 261.32 & -126.33 & 256.66 & 261.32 & 1.200 &       &       &       &         \\
2 & -107.97 & 223.94 & 233.26*& -108.02 & 222.04 & 229.03*& 1.131 & 0.099 & 0.753 & 0.610 & 0.0043  \\
3 & -107.94 & 227.87 & 241.86 & -108.02 & 224.05 & 233.37 & 1.126 & 0.172 & 0.917 & 0.933 & 0.0248  \\
4 & -103.96 & 223.92 & 242.57 & -104.20 & 218.40 & 230.05 & 1.087 & 0.476 & 0.924 & 0.902 & 0.0102  \\
\hline
\end{tabular}}
\caption{Turtle data (76 observations): Values of the NNTS parameter ($M$), log-likelihood, AIC, and BIC for general (columns 2-4) and symmetric NNTS (columns 5-7) models.
Estimated angle of reflection ($\hat{\mu}$) is in column 8, loglikelihood ratio test statistic ($LR_{GS}$) in column 9, with its asymptotic chi-squared and bootstrapped ($BLR$)
p-values in columns 10 and 11. Column 12 contains the Wald-type symmetry measure ($SK_{NNTS}$). 
\label{turtlesloglikAICBIC}}
\end{center}
\end{table}

\begin{table}[t]
\begin{center}
\scalebox{0.6}{
\begin{tabular}{c||ccc||ccc||c||ccc||cccc}
\hline
     & \multicolumn{3}{|c||}{General NNTS model} & \multicolumn{3}{c||}{Symmetric NNTS model} &              & \multicolumn{3}{c||}{Likelihood Ratio}     &    \\
$M$  & loglik & AIC & BIC                        & loglik & AIC & BIC                         & $\hat{\mu}$  & $LR_{GS}$ & $\chi^2$ p-value & BLR p-value & $SK_{NNTS}$ \\
\hline
\hline
0 & -393.31 & 786.61 & 786.61 &         &        &        &       &        &       &       &       \\
1 & -382.76 & 769.53 & 776.26 & -382.76 & 769.53 & 776.26 & 6.279 &        &       &       &       \\  
2 & -294.56 & 597.12 & 610.58 & -299.14 & 604.28 & 614.38 & 1.556 &  9.161 & 0.002 & 0.002 & 0.0306 \\  
3 & -285.83 & 583.67 & 603.86 & -307.66 & 623.31 & 636.78 & 1.544 & 43.646 & 0.000 & 0.001 & 0.2605 \\  
4 & -258.02 & 532.04 & 558.97 & -262.71 & 535.41 & 552.24*& 1.515 &  9.371 & 0.025 & 0.084 & 0.0846 \\  
5 & -248.71 & 517.43 & 551.09*& -271.39 & 554.77 & 574.97 & 4.682 & 45.345 & 0.000 & 0.001 & 0.3467 \\  
6 & -244.54 & 513.08 & 553.47 & -264.18 & 542.36 & 565.92 & 4.679 & 39.277 & 0.000 & 0.016 & 0.1893 \\  
7 & -241.45 & 510.90 & 558.02 & -272.22 & 560.43 & 587.36 & 4.681 & 61.534 & 0.000 & 0.001 & 0.3159 \\  
8 & -240.81 & 513.63 & 567.48 & -269.96 & 557.92 & 588.21 & 4.673 & 58.291 & 0.000 & 0.001 & 0.2979 \\  
\hline
\end{tabular}}
\caption{Dragonfly data (214 observations): Values of the NNTS parameter ($M$), log-likelihood, AIC, and BIC for general (columns 2-4) and symmetric NNTS (columns 5-7) models.
Estimated angle of reflection ($\hat{\mu}$) is in column 8, loglikelihood ratio test statistic ($LR_{GS}$) in column 9, with its asymptotic chi-squared and bootstrapped ($BLR$)
p-values in columns 10 and 11. Column 12 contains the Wald-type symmetry measure ($SK_{NNTS}$). 
\label{dragonflyloglikAICBIC}}
\end{center}
\end{table}

\begin{table}[t]
\begin{center}
\scalebox{0.6}{
\begin{tabular}{c||ccc||ccc||c||ccc||cccc}
\hline
     & \multicolumn{3}{|c||}{General NNTS model} & \multicolumn{3}{c||}{Symmetric NNTS model} &              & \multicolumn{3}{c||}{Likelihood Ratio}     &    \\
$M$  & loglik & AIC & BIC                        & loglik & AIC & BIC                         & $\hat{\mu}$  & $LR_{GS}$ & $\chi^2$ p-value & BLR p-value & $SK_{NNTS}$ \\
\hline
\hline
0  & -569.74 & 1141.48 & 1145.22 &         &        &        &       &        &       &       &        \\
1  & -455.22 &  914.45 &  921.92 & -455.22 & 914.45 & 921.92 & 0.584 &        &       &       &        \\
2  & -409.66 &  827.32 &  842.27 & -422.89 & 851.77 & 862.98 & 0.327 & 27.078 & 0.000 & 0.001 & 0.0587 \\
3  & -391.68 &  795.36 &  817.78 & -405.54 & 819.07 & 834.02 & 0.157 & 27.764 & 0.000 & 0.001 & 0.0506 \\
4  & -373.95 &  763.90 &  793.79*& -392.11 & 794.22 & 812.91 & 0.201 & 36.754 & 0.000 & 0.001 & 0.0780 \\
5  & -370.98 &  761.96 &  799.33 & -386.32 & 784.63 & 807.05*& 0.188 & 31.215 & 0.000 & 0.001 & 0.0786 \\
6  & -366.67 &  757.34 &  802.18 & -383.93 & 781.85 & 808.01 & 0.226 & 35.407 & 0.000 & 0.001 & 0.0765 \\
7  & -362.12 &  752.24 &  804.56 & -384.45 & 784.90 & 814.79 & 0.163 & 45.335 & 0.000 & 0.001 & 0.1031 \\
8  & -356.68 &  745.36 &  805.14 & -382.94 & 783.89 & 817.52 & 0.207 & 60.537 & 0.000 & 0.001 & 0.1602 \\
9  & -354.66 &  745.33 &  812.59 & -379.17 & 778.35 & 815.71 & 0.138 & 50.637 & 0.000 & 0.001 & 0.1376 \\
10 & -353.21 &  746.42 &  821.15 & -379.99 & 781.97 & 823.07 & 0.195 & 64.782 & 0.000 & 0.001 & 0.1579 \\
11 & -351.95 &  747.90 &  830.11 & -378.07 & 780.13 & 824.97 & 0.151 & 55.090 & 0.000 & 0.001 & 0.1506 \\
12 & -351.38 &  750.75 &  840.43 & -381.25 & 788.49 & 837.07 & 0.138 & 62.626 & 0.000 & 0.001 & 0.1462 \\
\hline
\end{tabular}}
\caption{Wind data (310 observations): Values of the NNTS parameter ($M$), log-likelihood, AIC, and BIC for general (columns 2-4) and symmetric NNTS (columns 5-7) models.
Estimated angle of reflection ($\hat{\mu}$) is in column 8, loglikelihood ratio test statistic ($LR_{GS}$) in column 9, with its asymptotic chi-squared and bootstrapped ($BLR$)
p-values in columns 10 and 11. Column 12 contains the Wald-type symmetry measure ($SK_{NNTS}$). 
\label{windloglikAICBIC}}
\end{center}
\end{table}

\newpage

\end{document}